\documentclass[final,5p,times,twocolumn,authoryear]{elsarticle}

\usepackage{amssymb}
\usepackage{lipsum}
\usepackage{gensymb}
\usepackage{indentfirst} 
\usepackage{enumerate}
\usepackage{subcaption}
\usepackage{amsmath}
\usepackage{booktabs}
\usepackage{bm}
\usepackage{xcolor}
\usepackage{natbib}
\usepackage[colorlinks=true,linkcolor=blue,citecolor=blue,urlcolor=blue]{hyperref}
\usepackage{cancel}
\journal{High Energy Astrophysics}

\begin{document}

\begin{frontmatter}



\title{Numerical studies of (in)stabilities of shocks in perturbed advective flows around black holes}


\author[first]{Junxing Zhou
\author{Junxing Zhou 
\corref{cor2}\fnref{label2}}
\ead{zhoujunxing@mail.ynu.edu.cn}}
\author[first]{Junxiang Huang}
\author[first]{Xin Chang}
\author[second]{Toru Okuda}
\author[first]{Chandra B. Singh 
\author{Chandra B. Singh \corref{cor1}\fnref{label1}}
\ead{chandrasingh@ynu.edu.cn}}
\affiliation[first]{organization={South-Western Institute for Astronomy Research, Yunnan University},%Department and Organization
            addressline={Chenggong District}, 
            city={Kunming},
            postcode={650500}, 
            state={Yunnan},
            country={China}}
            
\affiliation[second]{organization={Hakodate Campus, Hokkaido University of Education},
addressline={Hachiman-Cho 1-2},
city={Hakodate}, postcode={040-8567}, state={Hokkaido}, country={Japan}
}
\begin{abstract}
%The accretion flow is one of the main components of high-energy astrophysical phenomena around black holes. 
Using two-dimensional hydrodynamic simulations, we investigate the stability of shocked accretion flows around black holes under non-axisymmetric perturbations. By systematically exploring the parameter space of specific energy and angular momentum that permits shock formation in advective accretion flows, we demonstrate that quasi-periodic oscillations (QPOs) naturally emerge in perturbed systems. Our spectral analysis reveals characteristic QPO frequencies spanning 0.44-146.57$Hz$, effectively bridging the observed low-frequency (LFQPOs) and high-frequency QPOs (HFQPOs) in black hole X-ray binaries. The quality factors ($Q = \mu_0/2\Delta$) of these oscillations range from 1.66 to 203.58, with multiple Lorentzian components indicating distinct oscillation modes. Through wavelet analysis and cross-validation with recent observations (e.g., Swift J1727.8-1613 and GX 339-4), we establish that shock instabilities driven by acoustic wave interactions between the non-axisymmetric perturbation and the shock location can quantitatively explain the temporal features observed in accreting black hole systems. Furthermore, we characterize the $\gamma$-dependence of shock morphology, showing that increasing the adiabatic index from 4/3 to 1.4 changes shock positions outward while maintaining oscillation coherence.
\end{abstract}



\begin{keyword}
%% keywords here, in the form: keyword \sep keyword, up to a maximum of 6 keywords
Accretion disks \sep Black hole physics \sep QPOs \sep Numerical simulation



\end{keyword}

\end{frontmatter}

%\tableofcontents



\section{Introduction}
\label{introduction}
Astrophysical black holes (BHs) are one of the simplest compact objects with only two basic parameters, mass and spin. They vary in a wide range of masses, stellar-mass ($3 - 10^{2} M_{\odot}$), intermediate mass ($10^{2} - 10^{5} M_{\odot}$) and supermassive ($10^{5} - 10^{9} M_{\odot}$). Nothing escapes from inside the horizons of the BHs, so there is only an indirect method to imply about their properties.  The emission properties of accretion flows around BHs are one of the observational signatures whose properties are correlated with those of black holes.

%Black holes (BHs) as one of the types of compact objects have drawn widespread interest since 2019 when the Event Horizon Telescope (EHT) collaborations presented several results of the shadow of the compact object in Sagittarius A* (or Sgr A*) and M87* \citep{2019ApJ...875L...1E, 2021ApJ...910L..13E, 2022ApJ...930L..12E}. The flow dynamics suggested that hot accretion flows surround these BHs. BHs are compact objects widely accepted on the basis of observational evidence. However, those are also bizarre objects; even light cannot escape from their horizons. The properties of black holes are implied by the study of accretion flows.

The systems containing stellar-mass black holes and a companion star are known as black hole X-ray binaries (BHXRBs). In such systems, the matter is supplied from the star to the black hole and emission properties of the accretion flows are observed in the range of wavelengths, especially X-rays close to the black holes. Some BHXRBs show transient nature and the corresponding changes in spectral states are seen when they transition between states like quiescence and outbursts. There are basic spectral states, namely, low hard (LHS), high soft (HSS) and intermediate (IMS) (\citealt{2006ARA&A..44...49R} and references therein). When BHXRBs usually change spectral states during their evolution processes, they also exhibit corresponding temporal features, namely quasi-periodic oscillations (QPOs). QPOs in BHXRBs are classified into two types, low-frequency QPOs (LFQPOs) with centroid frequency $\le$ 30 Hz and high-frequency QPOs (HFQPOs) with centroid frequency $\ge$ 60 Hz (\citealt{2010LNP...794...53B} and references therein). Several models have been proposed to explain the physical scenarios for various types of QPOs around BHXRBs.(\citealt{2019NewAR..8501524I} and references therein). 
%Such temporal properties are likely to be associated with spectral states of accretion flows around the BHs. 

A physically realistic accretion model should have the potential to address the origin of spectral as well as temporal properties of emissions around black holes. Since the 1950s, accretion flows around BHs have been extensively studied. Initially, the most basic picture was explored. The flow was assumed spherically symmetric and with zero angular momentum known as the Bondi accretion flow which is a radiatively inefficient flow\citep{1952MNRAS.112..195B}. Another class of accretion flow was proposed by Shakura \& Sunyaev with a Keplerian profile and became a famous fundamental model called the standard cold thin accretion disk (SSD) model which is a radiatively efficient flow \citep{1973A&A....24..337S}.  In subsequent years, interest of the community shifted to the study of the hot accretion flow, one of the types is advection-dominated accretion flow (ADAF), where most of the energy generated by viscous dissipation is stored as entropy rather than being radiated away \citep{1994ApJ...428L..13N}. ADAF model with a single sonic point demonstrated its ability to better represent the observed features of BHXRBs and supermassive black holes (\citealt{2014ARA&A..52..529Y} and references therein). Besides the above-mentioned works, generalized advective accretion flows with two components and multiple sonic points have been explored in detail since the 1990s. A dynamically active, geometrically thick, and optically thin hot accretion flow coexisting with the geometrically thin and optically thick cold flow is instrumental in interpreting the spectral as well as temporal changes in radiations from various X-ray sources with black holes \citep{2000ApJ...531L..41C, 2021ApJ...920...41M}. The properties of the corona are self-consistently elucidated by a theoretical solution of a hybrid flow that undergoes a standing shock, manifesting as post-shock sub-Keplerian accreting matter, potentially giving rise to outflows and hard X-rays \citep{1995ApJ...455..623C}. 

All the above-mentioned accretion flow models are derived from solving a set of conservation equations. However, it is essential to examine whether these solutions can still represent realistic scenarios when perturbations are present.\citep{1998pfp..book.....C}. To study the stability properties of shocks in the advective flows around BHs, \citet{1999ApJ...516..411M} performed two-dimensional (2D) smoothed particle hydrodynamics (SPH) simulations using the pseudo-Newtonian potential for a non-rotating BH. In their work, they found QPOs in the perturbed systems and these can have a relation with some relevant observational signatures. The mechanism of this instability has been proposed to be associated with the cycle of acoustic waves between the corotation radius and the shock (\citealt{2006MNRAS.365..647G}, and reference therein). Further works with general relativistic treatment were done, and the perturbed shock waves were studied for nonrotating as well as rotating BHs \citep{2008ApJ...689..391N,2009ApJ...696.2026N}.
%The shocked accretion flow was also examined taking into account heating and cooling processes. In the presence of Compton cooling, the thermal pressure within the corona decreases, leading the shock to move closer to the central object 
% \citep{1996ApJ...470..460M}. Conversely, the outward transportation of angular momentum due to viscosity causes the shock to shift outward, making the angular momentum of the accretion flow approach the Keplerian values at a large distance \citep{1995MNRAS.272...80C}. These simulations were carried out in two dimensions (2D), based on axisymmetric assumptions.
Recently, a three-dimensional (3D) simulation of sub-Keplerian matter accretion was performed using Eulerian hydrodynamic code and it established that it is possible to generate a shock surface supported by the centrifugal barrier \citep{2023MNRAS.519.4550G}. They also explored a few cases of perturbed standing shocks and presented the physical scenario.
%QPOs are most effectively analyzed using the fast Fourier transform. They appear as sharp peaks in the power spectrum, derived from the squared magnitude of the fast Fourier transform of the light curve. These peaks have a width typically less than half of their central frequency \citep{2019NewAR..8501524I}. 
Moreover, oscillating shocks in low angular momentum accretion flows can cause variations in the luminosity of black holes which can be deemed as QPOs, and for specific parameters, the oscillations of the shock position correspond to observed variations in black hole luminosity (\citealt{2015MNRAS.447.1565S} and reference therein). The global shock oscillations caused by SASI (Standing Accretion Shock Instability) have been proposed as a potential mechanism to explain the origin of QPOs in case of the Bondi flows around BHs \citep{2018MNRAS.476.3310D}. Specifically, the shock oscillations resulting from SASI can lead to observable periodic variations in the light curve, which correspond to the QPOs observed in the systems containing compact objects. In the post-shock region, the compression of gases generates outward-propagating acoustic waves and inward-propagating entropy-vorticity waves. The interaction between these two types of waves forms a cycle that leads to the instability of the shock. Over time, this instability grows exponentially, eventually leading the system to reach a quasi-steady state characterized by stable nonlinear oscillations.

Our work presents the 2D simulations of the advective flows with shocks using high-resolution shock-capturing grid-based, finite volume computational fluid dynamics, Eulerian code PLUTO. We explore the whole parameter space of specific energy and specific angular momentum of the accretion flow which can allow shock transitions and study the consequences of perturbing such shocked flows around the BHs. The organization of this paper is as follows: In Section 2, we describe the analytical approach, including how to calculate the sonic points and the shock location in the flows around BHs. Section 3 contains the details of the computational domain, initial and boundary conditions and the results of our simulations, along with our analyses and interpretations of these results. In Section 4, we relate our numerical findings with the current observational signatures. Finally, Section 5 contains the summary and conclusions.

\section{Theoretical Solution}
%%\label{}
We consider the transonic, inviscid,  adiabatic low angular momentum accretion flows around BHs. We have the continuity equation for mass conservation and the Bernoulli constant representing specific energy conservation to describe dynamics \citep{1989ApJ...347..365C,1992MNRAS.259..410C,1993ApJ...417..671C},

\begin{equation}
\dot{M} = \textbf{v}_r\rho r h,
\label{eqn:accretionrate}
\end{equation}
\begin{equation}
B = \frac{1}{2}\textbf{v}_r^2 +\frac{C_s^2}{\gamma-1}+\frac{\lambda^2}{2r^2}+g(r),
\label{eqn:Bernoulli}
\end{equation}
where the $\dot{M}$ is the accretion rate, $B$ is the Bernoulli constant, $h$ is the constant height of the flow, $\lambda$ is the specific angular momentum, $\gamma$ is the adiabatic index, and $g(r)$ is the gravitational potential. $C_s$ and $\textbf{v}_r$ are sound speed and radial velocity in units of speed of light respectively. The length and time scales are taken in units of Scharzschild radius $r_{g} = 2G\mathbb{M}/c^2$ and $2GM/c^3$ respectively, $G$ and $\mathbb{M}$ being the gravitational constant and black hole mass respectively. We consider the pseudo-Newtonian potential $g(r)=-1/2(r-1)$\citep{1980A&A....88...23P} to simulate the general relativistic effect around a Schwarzschild BH. The relation between gas pressure and density is as follows,
\begin{equation}
p =\frac{\rho C_s^2}{\gamma}
\label{eqn:pressure}.
\end{equation}

\begin{figure}
    \centering
    \includegraphics[width=.5\linewidth]{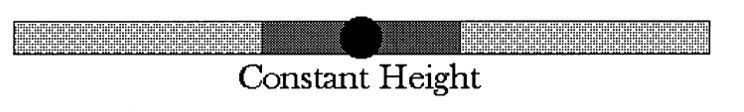}
    \caption{Cartoon diagram for constant height model, also known as model H, adapted from \citet{2001MNRAS.327..808C}}
    \label{fig：constantheight}
\end{figure}

In this work, the flow motion is restricted in the  $XY$ plane only with no vertical motion, the height of the flow is a constant, so the accretion flow structure can be represented as Fig.\ref{fig：constantheight} (\citealt{2001MNRAS.327..808C}, hereafter to be referred as \textbf{CD01}).
 
For $\gamma=4/3$, the energy equation is transformed into a quartic equation as,
\begin{equation}
    4Br^4 - (8B+5)r^3 + (4B+12\lambda^2-2)r^2 - 24\lambda^2r+12\lambda^2=0.
    \label{eqn:criticalpoint}
\end{equation}

There is only one unknown $r$, while fixed values of $B$ and $\lambda$ are taken as free parameters for the accretion flow. Solving the set of conservation equations yields four values for $r$, these values represent the critical points. Among these, one point is located inside the event horizon, and we are interested in the other critical points, namely inner and outer ones. Following the solution procedure described in \textbf{CD01}, we obtain a gradient of the radial velocity.
\begin{equation}
\frac{d\mathbf{v}_r}{dr}=\frac{\frac{C_s^2}{r}\mathbf{v}_r+\frac{\lambda^2\mathbf{v}_r}{r^3}-\frac{\mathbf{v_r}}{2(r-1)^2}}{\mathbf{v}_r^2-C_s^2}.
    \label{eqn:vrrelation}
\end{equation}
%To solve this differential equation, the two sonic points are obtained from Eq. \ref{eqn:criticalpoint}.%
Once the specific energy and angular momentum are determined, we find the sonic points using Eq. \ref{eqn:criticalpoint}. At the sonic point, the velocity of the accretion flow equals the sound speed, so we set \( C_s = \mathbf{v_r} \) in Eq. \ref{eqn:Bernoulli}, and the only unknown is \( \mathbf{v_r} \). For this differential equation, we took the odeint method to get the function curves. The odeint is a numerical integration method used for solving ordinary differential equations (ODEs)\citep{article}. It's based on the Runge-Kutta method, specifically the fourth-order Runge-Kutta scheme, which is widely used for its balance between accuracy and computational efficiency. It's an explicit method, meaning it doesn't require solving implicit equations to update the solution and it always includes adaptive step size control, which adjusts the step size based on the dynamics of the solution to improve accuracy and efficiency.

\begin{figure}
    \centering
    \includegraphics[width=1\linewidth]{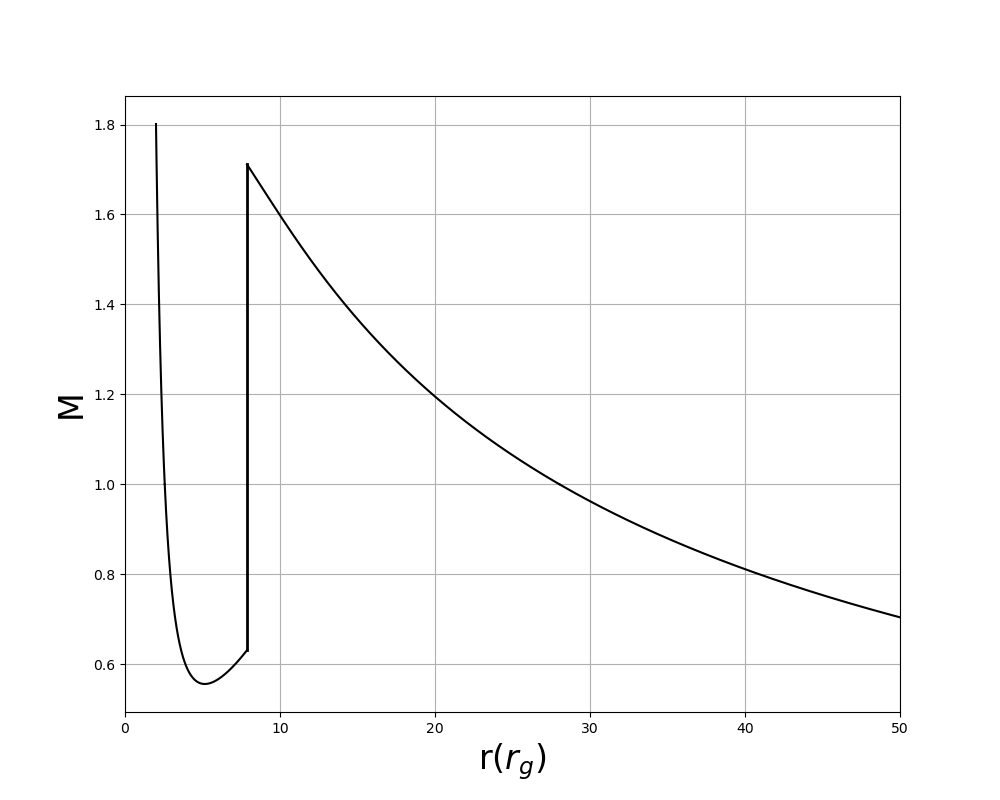}
    \caption{Mach number versus radial distance. The shock location is at the 7.87$r_g$.}
    \label{fig:shocklocation}
\end{figure}

Following the solution procedure in the \textbf{CD01}, we obtain the shock condition based on the energy flux conservation equation, the pressure balance condition and the baryon flux conservation equation,

\begin{equation}
    8M_+^2 M_-^2 - (M_+^2 + M_-^2) - 6 = 0.
    \label{eqn:shocklocation}
\end{equation}
Where $M_-$ and $M_+$ respectively represent the mach number of accretion flow before and after the shock location. We obtain the shock location located between the two sonic points as shown in Fig.\ref{fig:shocklocation}. In this example, the specific angular momentum $\lambda$ is 1.8 in units of $2GM/c$ and the specific energy $B$ is 0.03628 units of $c^2$\citep{2023MNRAS.519.4550G}, two sonic points are 28.02$r_g$ and 2.56$r_g$ respectively, and the shock location is 7.87$r_g$.

%\subsection{Subsection title}

\section{Numerical setup}
%%\label{}

%\subsection{Subsection title}
\label{subsec:simulation}
We employed the publicly available package PLUTO Code \citep{2007ApJS..170..228M} for our simulation work. We used the hydrodynamic (HD) module of the PLUTO code. This module can solve the Euler or Navier-Stokes equations of classical fluid dynamics. Using the HD module, PLUTO evolves according to the system of conservation laws
\[
\frac{\partial}{\partial t} \begin{pmatrix} \rho\\ m\\ E_t + \rho \Phi \end{pmatrix} + \nabla \cdot \begin{pmatrix} \rho \mathbf{v}\\ m \mathbf{v} + p I\\ \left( E_t + p + \rho \Phi \right) \mathbf{v} \end{pmatrix}^T = \begin{pmatrix} 0 \\ -\rho \nabla \Phi + \rho g \\m \cdot g \end{pmatrix}.
\]
Here, \(\rho\) denotes the mass density. The momentum density, represented by \(m\), is defined as the product of \(\rho\) and the velocity \(\mathbf{v}\). $ \Phi$ and $g$ represent gravitational potential and acceleration vector respectively.The thermal pressure is indicated by \(p\), and \(E_t\) signifies the total energy density:
\begin{equation}
    E_t=\rho e+\frac{m^2}{2\rho}.
\end{equation}
$e$ denotes the internal energy.
Original variables are often more straightforward and favoured for setting initial and boundary conditions and for use in interpolation processes. The vector of original variables, denoted as \( V \), follows the quasilinear form of the governing equations:
\begin{align}
    &\frac{\partial\rho}{\partial t}+\mathbf{v}\cdot\nabla\rho+\rho \nabla \cdot \mathbf{v}=0, \\
    &\frac{\partial \mathbf{v}}{\partial t}+\mathbf{v}\cdot \nabla \mathbf{v} + \frac{\nabla p}{\rho}=-\nabla \Phi + g, \\
    &\frac{\partial p}{\partial t}+\mathbf{v}\cdot \nabla p +\rho C_s^2\nabla \cdot \mathbf{v} =0.
\end{align}
The adiabatic speed of sound for an ideal equation of state (EOS) is given in Eq.\ref{eqn:pressure}.

%For testing the PLUTO code, we simulated a sample case of the Bondi accretion. We considered the sound speed value as $\sqrt{1/150}$ using the theoretical solution. We considered the natural units (i.e, $G = M = c =1$). The critical radius can be calculated \citep{1998bhad.conf.....K} by
%\begin{equation}
%r_c=\frac{GM}{2C_s ^2}.
%\end{equation}

%For the boundary condition in this case, the Mach number of fluid is $1.5$ at $50r_g$. The density value of the injected flow from outer boundary is taken as an arbitrary parameter so we set $\rho = 1$, $\gamma$ is 4/3 and the pressure can be calculated using Eq. \ref{eqn:pressure}. The initial background density on the XY plane is set as $1\times10^{-4}$. The coordinate system is spherical. We use the HLL Riemann solver, second-order-in-space linear interpolation, and the second-order-in-time Runge–Kutta algorithm. The computational domain in spherical co-ordinates is $2r_g \leq r \leq 50r_g, \quad 0 \leq \phi \leq 2\pi$, the grid is static and the number of physical cells are ($n_r$, $n_\phi$) = (100, 128).

%\begin{figure}[h]
%    \centering
%    \includegraphics[width=.8\linewidth ]{figure3.png}
%    \caption{Density of Bondi accretion flow at $1000 r_{g}/c$.}
%    \label{fig:simulationrm}
%\end{figure}

%The density distribution of the Bondi accretion model is shown in Fig. \ref{fig:simulationrm}. This simple simulation examines the accuracy of both the analytical solution and the numerical result from simulation run, the results are in good agreement.

\begin{figure}[h]
    \centering
    \includegraphics[width=.5\linewidth ]{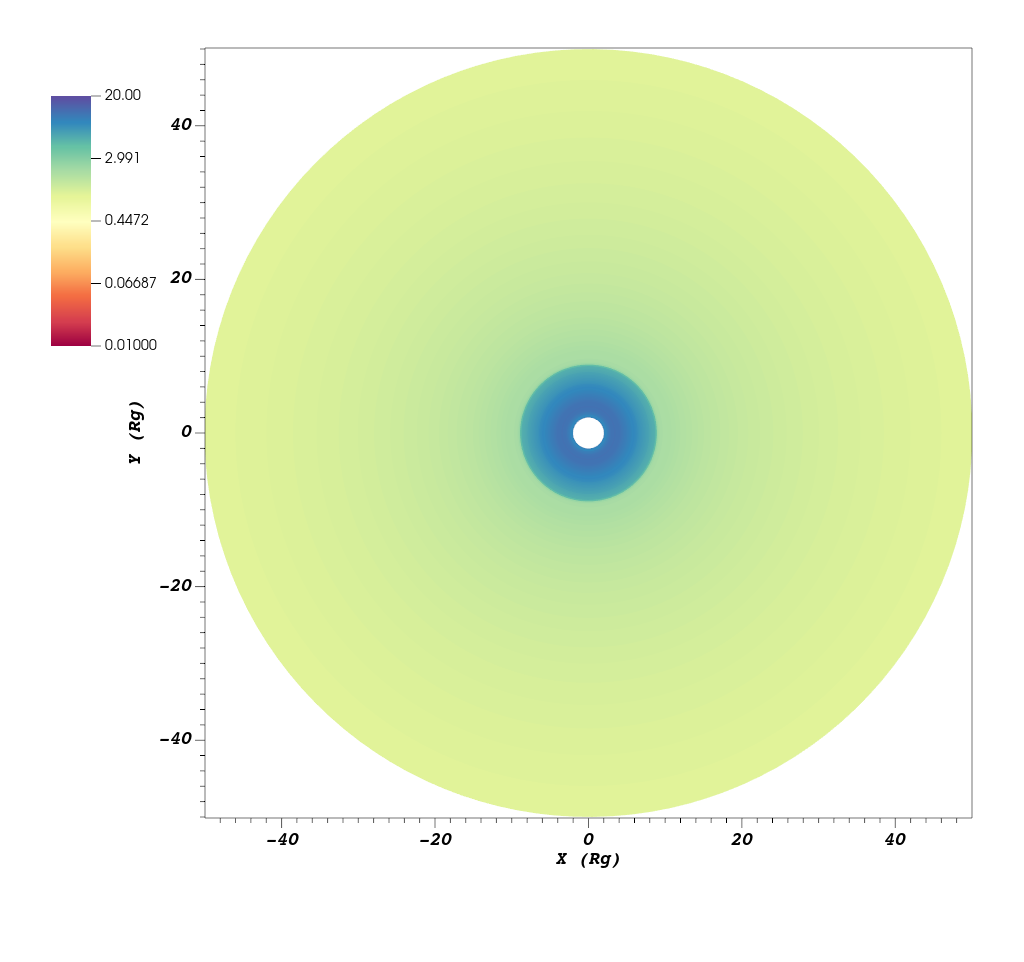}\hfill
    \includegraphics[width=.5\linewidth ]{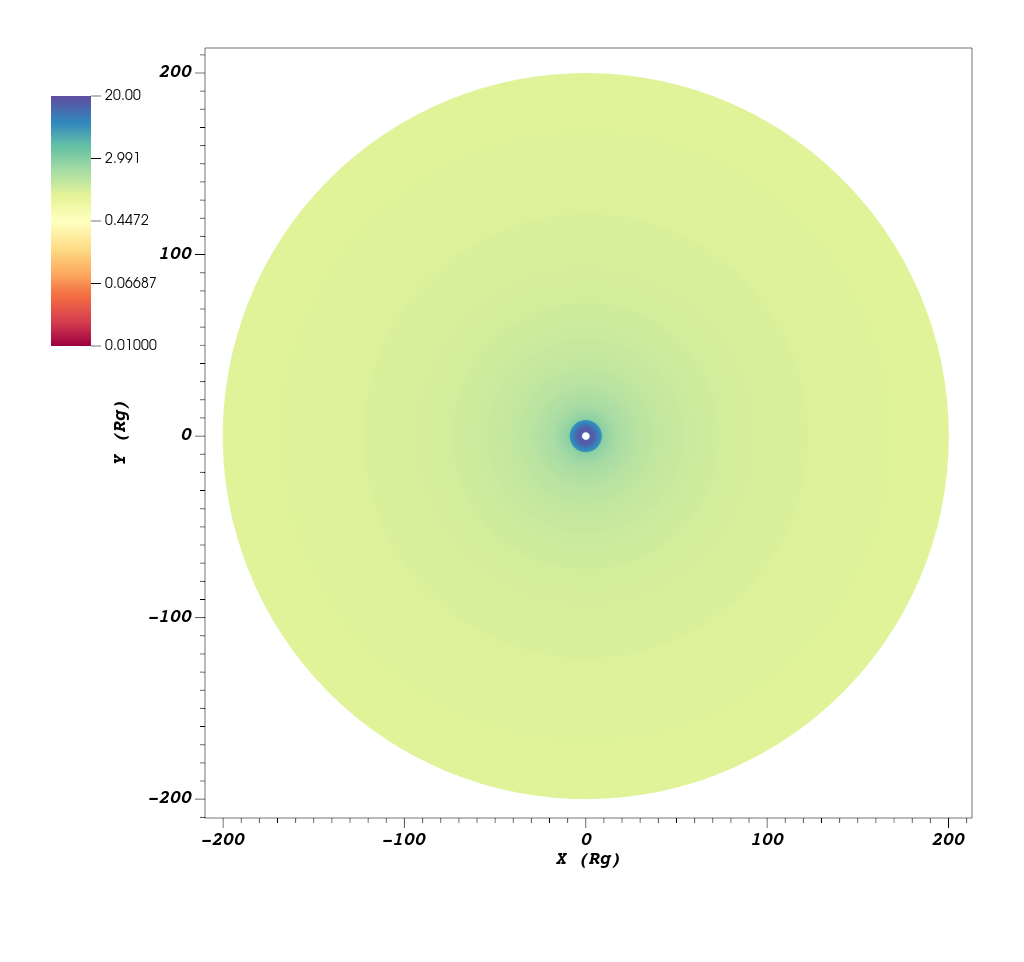}
    \caption{Cases with the same specific energy and specific angular momentum with computational domains of two sizes. The outer boundary in the left panel is 50$r_g$, while it is at 200$r_g$ in the right panel. }
    \label{fig:c3dm}
\end{figure}

For simulating the accretion flow with low angular momentum, we considered the cases with suitable specific angular momentum and the Bernoulli constant which allow the formation of the stable shock \citep{2006PhDT........11T}. We use the analytical results to initiate the simulation runs with different setups. We set the initial background density on the XOY plane as $1\times10^{-2}$ to avoid the negative density in the computational domain, and the density of injection flow at the outer boundary is 1. The black hole mass $\mathbb{M}$ is \(10 M_\odot\). The initial conditions are calculated as follows: with all the physical quantities at the sonic point known, we substitute them into Eq. \ref{eqn:vrrelation} and set the outer boundary, at which we can obtain the velocity and sound speed of the accretion flow. Since the value of density is arbitrary, we can obtain the pressure using Eq. \ref{eqn:pressure}. We use the HLL Riemann solver, second-order-in-space linear interpolation, and the second-order-in-time Runge–Kutta algorithm. The computing domain is expanded $2r_g \leq r \leq 200r_g, \quad 0 \leq \phi \leq 2\pi$, the grid is static and the number of physical cells is ($n_r$, $n_\phi$) = (1000, 125) in spherical coordinates. The end time of the simulation run is 150000$r_g/c$. We used the same initial specific energy and angular momentum with two different computing domains, specifically 50$r_g$ and 200$r_g$. As shown in Fig. \ref{fig:c3dm}, the shock location is unchanged. The detailed parameters are listed in Table\ref{table:parameter}. Here, $\lambda$ is the specific angular momentum, $B$ is the Bernoulli constant, and $r_s$ is the shock location. 

\begin{table}
\footnotesize
  \centering
  \caption{Initial parameters of injection flow.}
  \renewcommand{\arraystretch}{1.5} % Increase row height
   \begin{tabular}{p{1cm} p{1cm} p{1cm} p{1cm} p{1cm} p{1cm} p{1cm}}
    \toprule
    Case  & $\lambda$   & $B$        & $r_s$ & $v_r$         & $C_s$ \\
    \midrule
    1     &   1.7770     & 0.04333    &  6.47 &  0.025950  &  0.1231080 \\
    \hline
    2     &   1.8000     & 0.04000    &  14.56 & 0.026860   &  0.1184780 \\
    \hline
    3     &   1.8100    & 0.03170     &  8.74  & 0.029810   &  0.1060300\\
    \hline
    4     &   1.8200     & 0.03332    &  15.91  &0.029130   & 0.1085770 \\
    \hline
    5     &   1.8571    & 0.02100    &  18.45  &0.034930   & 0.0872920 \\
    \hline
    6     &  1.8750      & 0.01707    &  24.98  &0.037595   & 0.0792373 \\
    \hline
    7     &   1.8900   & 0.01495    &  48.45  &0.039107   & 0.0745056  \\
    \hline
    8     &   1.9000    & 0.01095    & 35.21   &0.042839   & 0.0645494   \\
    \hline
    9     &   1.9188    & 0.00617    & 43.48   &0.049125   & 0.0497656   \\
    \bottomrule
  \end{tabular}
  \parbox{\textwidth}{\footnotesize \baselineskip 3.8mm}
\label{table:parameter}
\end{table}

We discuss case 3 as an example to illustrate the effects of non-axisymmetric perturbations on the shocked accretion flow. It is worth noting that the stability test of the shock is crucial for the simulation and associated physical relevance. As shown in Fig.\ref{fig:c3dm}, the accretion flow exhibits a stable shock. Furthermore, we slightly changed the density of a narrow region for an extremely short period. The details of introducing the perturbation are as follows: during simulation time $29970(r_g/c)\leq t\leq30000(r_g/c)$, we introduce a non-axisymmetric density perturbation of 1.01 in a narrow region of $30 r_g \leq r \leq 30.5 r_g, \quad 0 \leq \phi \leq 0.1$. When perturbation encounters the shocked region, the post-shock region becomes unstable. In other cases as well, we use the same method to introduce the perturbation, the location of the perturbation is different depending on the shock location. We introduced the perturbation closest possible to the shock location in all the simulation runs.\\ 

\begin{figure}[h]
    \centering
    \includegraphics[width=.5\linewidth ]{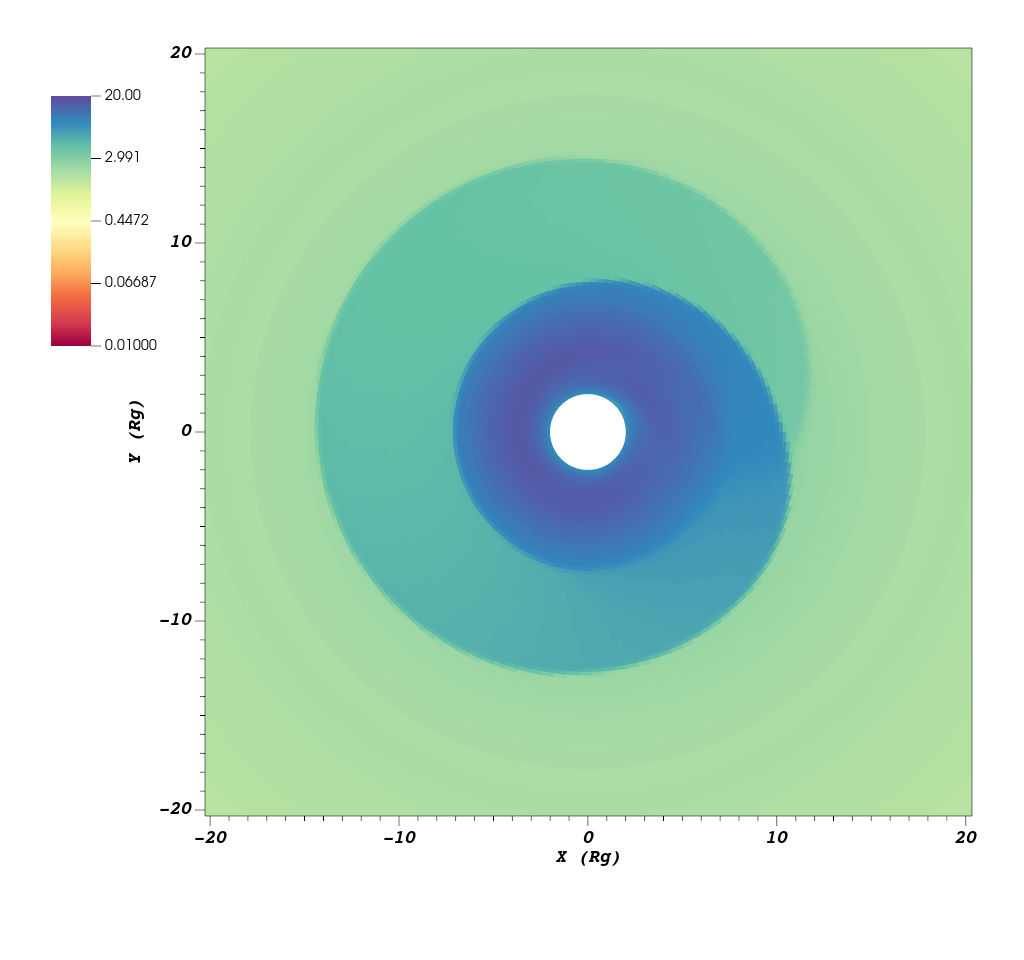}\hfill
    \includegraphics[width=.5\linewidth ]{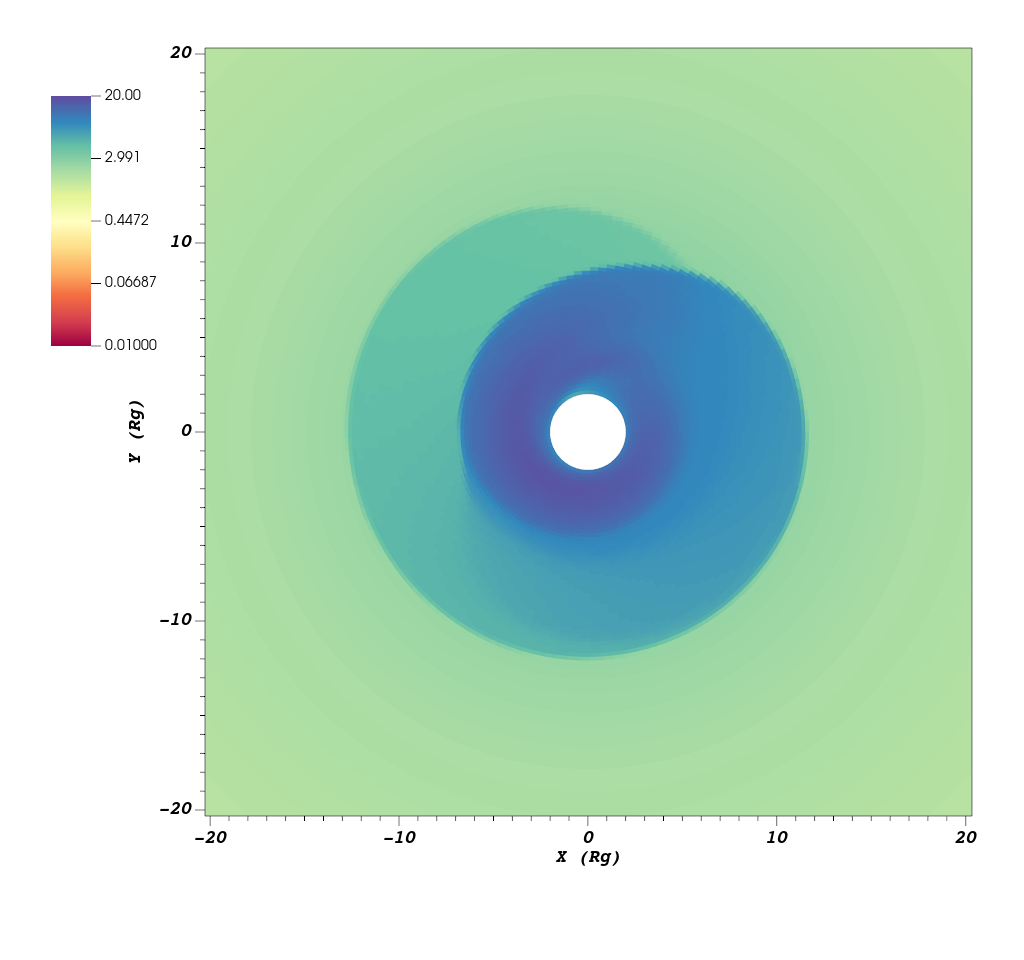}
    \caption{Evolution of density distribution for Case 3 at t = 50000$r_g/c$ \& 75000$r_g/c$.}
    \label{fig:c33}
\end{figure}

\section{Results}
\begin{figure}
    \centering
    \includegraphics[width=1\linewidth]{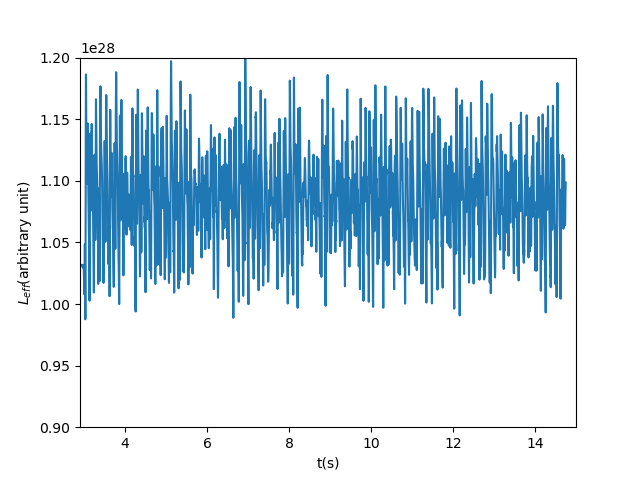}
    \caption{The light curve for case 3. The horizontal axis shows the simulation time in seconds, and the vertical axis shows the total luminosity in arbitrary units.}
    \label{fig:lightcurve}
\end{figure}

We used Shockfinder software \citep{Shockfinder} to analyze all the simulation data. In our present HD work, the proxy for radiative processes is the Bremsstrahlung. To explore QPOs, we focused on the light curve of the accretion disk. Firstly, we calculated the light curve, an example shown in Fig.\ref{fig:lightcurve}, however, it is not easy to extract the information about periodicity from the light curve. Based on the light curve, we performed different transformations. The light curve was estimated from the Bremsstrahlung emission, and the integrated total luminosity is as follows \citep{1979rpa..book.....R}
\begin{equation}
    L_{br}=g_{ff}\int1.4*10^{-27}(\frac{\rho}{m_p})^2 T^{\frac{1}{2}}\,dV.
\end{equation}
Here, $m_p$ is the proton mass, T is the temperature, and $g_{ff}$ is the Gaunt factor taken to be unity for simplicity. As shown in Fig.\ref{fig:lightcurve}, we calculate the Bremmstrahlung for case 3.  However, the periodicity of the light curve is not easy to distinguish so we performed a Fast Fourier Transform (FFT) of the light curve of the emission from the accretion flow after the perturbation was introduced, and the shock is unstable and oscillating. Of the 9 cases shown in Table \ref{table:parameter}, cases 1, 3, 5, 6, and 7 show quasi-periodicity. The QPO frequency distribution ranges from 0.49 to 146.57 Hz.

%FFT is an efficient algorithm for computing the Discrete Fourier Transform (DFT) and its inverse. It is a useful tool for detecting and quantifying periodic fluctuations in time series data, assuming the fluctuations have a truly constant period, amplitude, and phase. Even in time series with relatively consistent periodicity, we typically observe the evolution of fluctuation parameters over time. In this work, FFT converts the light curve from its time domain into the frequency domain. This transformation reveals the signal's frequency components, allowing for the analysis of periodic patterns and frequencies within the data. 

We performed multiple Lorentzian fits on the results of the FFT and identified significant QPO features. In the work led by \citet{2002ApJ...572..392B}, the value of Q (quality factor) is used to describe the sharpness of the Lorentzian function. The Q value is defined as the ratio of half of the central frequency ($\mu_0$) to the half-width at half maximum (HWHM, $\Delta$) of the Lorentzian, that is,
\begin{equation}
    Q = \frac{\mu_0}{2\Delta}.
\end{equation}
Here, $\mu_0$ is the central frequency of the Lorentzian, which is the position of the Lorentzian peak, and $\Delta$ is the half-width at half maximum of the Lorentzian, which is the width of the function when the value drops to half of its maximum. A higher Q value indicates a sharper Lorentzian, meaning that the frequency component is narrower, which is usually associated with QPOs of higher coherence. Conversely, a lower Q value indicates a broader Lorentzian, corresponding to noise components of lower coherence. The cases with QPOs and their Q values are listed in Table \ref{table:result}. Of the total cases, only two have Q value smaller than 2, whereas the remaining cases have Q value larger than 2, indicating that five cases show quasi-periodicity.

\begin{table}
\footnotesize
  \centering
  \caption{Frequency vs. Q values}
  \renewcommand{\arraystretch}{1.5} % Increase row height
   \begin{tabular}{p{1.2cm} p{1.2cm} p{1.2cm} p{1.2cm} p{1.2cm}}
    \toprule
    Case  & First peak($Hz$)   &Second peak($Hz$)    &Q for first peak  &Q for second peak \\
    \midrule
    1     &   40.17     & 146.57  &50.28   & 203.58  \\ 
    \hline
    3     &   8.15      & 28.89  & 7.69   &  9.26    \\
    \hline
    5     &   1.99      & 8.61   & 2.21   &  1.68     \\
    \hline
    6     &   1.65      & 8.99   & 3.75   &  1.66     \\
    \hline
    7     &   0.49      & 2.90   & 7.66   &  2.50      \\
    \bottomrule
  \end{tabular}
  \parbox{\textwidth}{\footnotesize \baselineskip 3.8mm}
\label{table:result}
\end{table}

While Fourier analysis can identify and to some extent quantify this non-stationary behaviour, it is not the most optimal approach for handling such complexities. To more accurately capture and describe the intricate temporal variations present in actual astrophysical data, we chose to employ more flexible and adaptive methods. Techniques like the wavelet transform, which provides localized information across different time scales, can better address non-constant periodicities and offer a more precise reflection of the true characteristics of the data. The wavelet transform holds potential for analyzing periods in time series, especially in identifying the time evolution of parameters such as period, amplitude, and phase in periodic and pseudoperiodic signals. By reformulating the wavelet transform as a projection, we can derive its statistical properties and develop improved rescaled transforms. Specifically, considering it as a weighted projection, the weighted wavelet Z-transform (WWZ) (\citealt{1996AJ....112.1709F}, \citealt{2017zndo....375648A}, \citealt{2024ApJ...976...51G} and reference therein), which enhances its capability to detect and quantify periodic and quasiperiodic signals more effectively.

The WWZ results demonstrate the frequency at which the signal exhibits the strongest periodicity. The deeper the red colour, the stronger the periodicity. We compared the results of the two analysis methods, FFT and WWZ as shown in the Figures below. Both FFT and WWZ exhibit a nice agreement in the frequencies of the QPOs. 

Out of a set of nine systematically conducted simulations, five cases exhibited detectable quasi-periodic oscillations (QPOs). Analysis of results summarized in Tables \ref{table:parameter} and \ref{table:result} demonstrate that shock positions are primarily governed by the flow's specific angular momentum and energy and smaller radial distance of the shock correlates with higher characteristic frequencies (40.17–146.67$Hz$ vs 0.44–8.99$Hz$). The observed WWZ-FFT discrepancies stem from methodological differences: FFT incorporates the full temporal evolution including initial transients (t=3.0–15.0$s$ across cases), while WWZ selectively enhances localized features through its adaptive wavelet basis, achieving superior time-frequency resolution at the expense of complete temporal coverage.

\begin{figure}[ht]
    \centering
    \begin{subfigure}[b]{0.45\textwidth}
        \includegraphics[width=\textwidth]{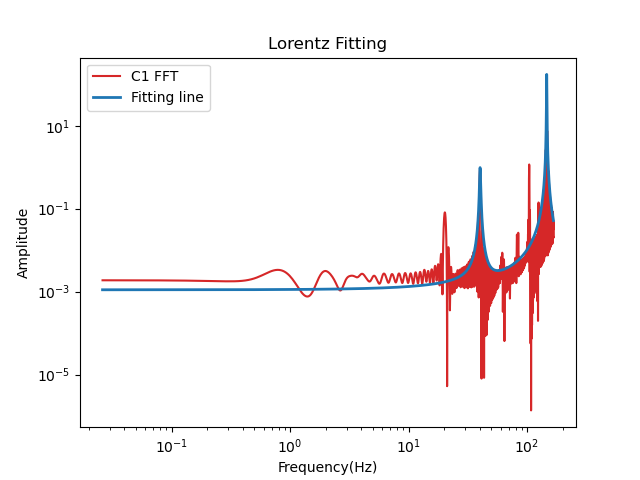}
        \caption{FFT of light curve}
        \label{fig:c1}
    \end{subfigure}
    \hfill
    \begin{subfigure}[b]{0.45\textwidth}
        \includegraphics[width=\textwidth]{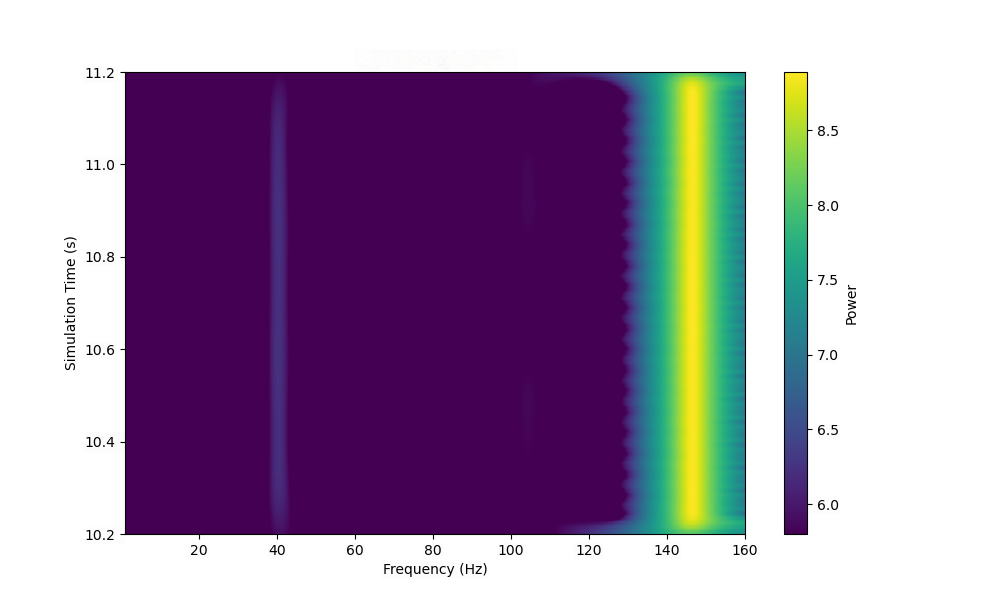}
        \caption{Weighted Wavelet Z-Transformation}
        \label{fig:c11w}
    \end{subfigure}
    \caption{Power density spectra of integrated total luminosity} for Case 1. Figure (a): the FFT result shows the QPOs frequency at 40.17$Hz$ and 146.57$Hz$. Figure (b): the WWZ result shows the strongest periodical frequency at 146.67$Hz$.
    \label{fig:c1two}
\end{figure}

\begin{figure}[ht]
    \centering
    \begin{subfigure}[b]{0.45\textwidth}
        \includegraphics[width=\textwidth]{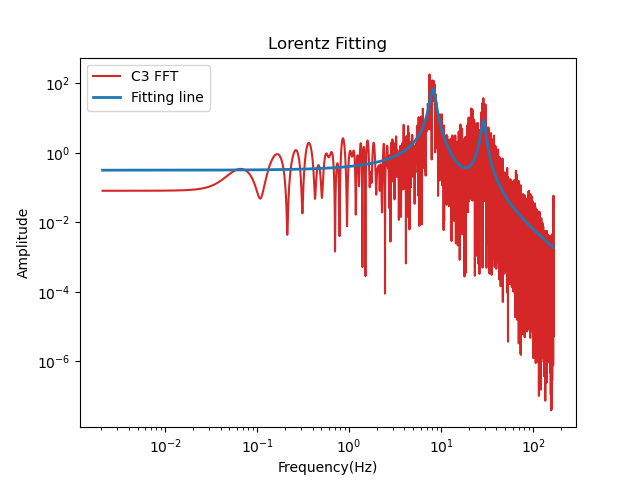}
        \caption{FFT of light curve}
        \label{fig:c3}
    \end{subfigure}
    \hfill
    \begin{subfigure}[b]{0.45\textwidth}
        \includegraphics[width=\textwidth]{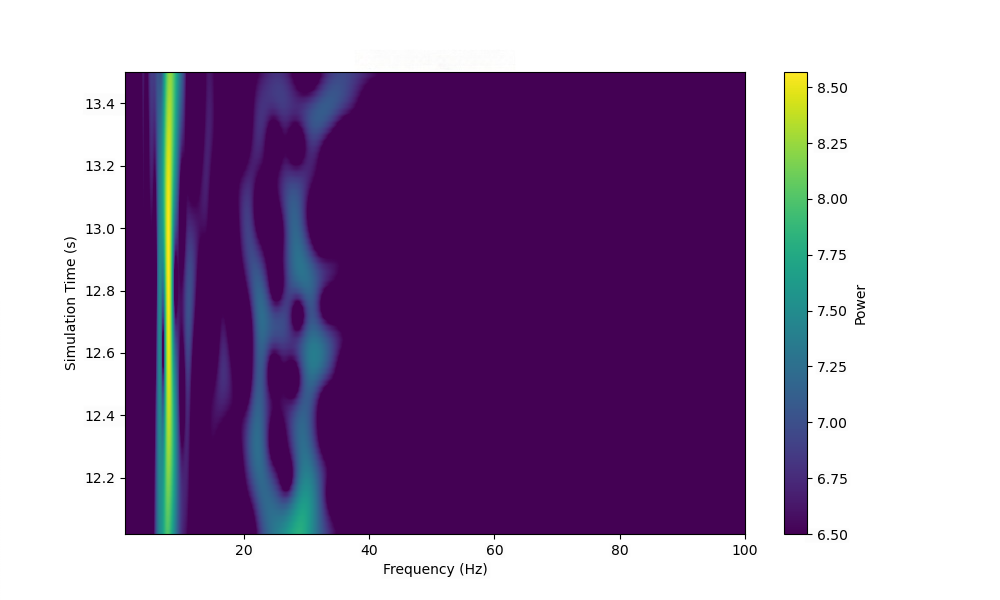}
        \caption{Weighted Wavelet Z-Transformation}
        \label{fig:c3w}
    \end{subfigure}
    \caption{Power density spectra of integrated total luminosity for Case 3. Figure (a): The FFT result shows the QPOs frequency are 8.15$Hz$ and 28.89$Hz$. For figure (b), the WWZ result shows the strongest periodical frequency is 7.85$Hz$.}
    \label{fig:c3two}
\end{figure}

\begin{figure}[ht]
    \centering
    \begin{subfigure}[b]{0.45\textwidth}
        \includegraphics[width=\textwidth]{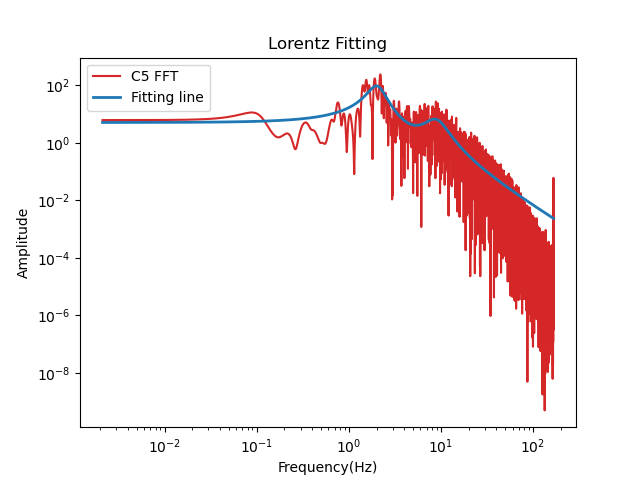}
        \caption{FFT of light curve}
        \label{fig:c5}
    \end{subfigure}
    \hfill
    \begin{subfigure}[b]{0.45\textwidth}
        \includegraphics[width=\textwidth]{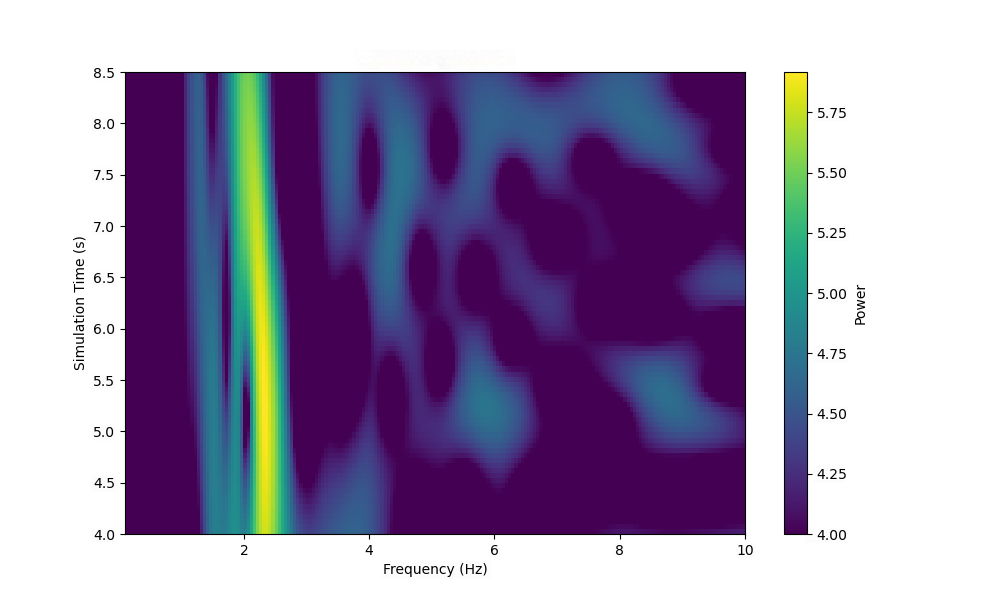}
        \caption{Weighted Wavelet Z-Transformation}
        \label{fig:c5w}
    \end{subfigure}
    \caption{Power density spectra of integrated total luminosity for Case 5. Figure (a): The FFT result shows the QPOs frequency at 1.99$Hz$ and 8.61$Hz$. Figure (b): The WWZ result shows the strongest periodical frequency at 2.28$Hz$.}
    \label{fig:c5two}
\end{figure}

\begin{figure}[ht]
    \centering
    \begin{subfigure}[b]{0.45\textwidth}
        \includegraphics[width=\textwidth]{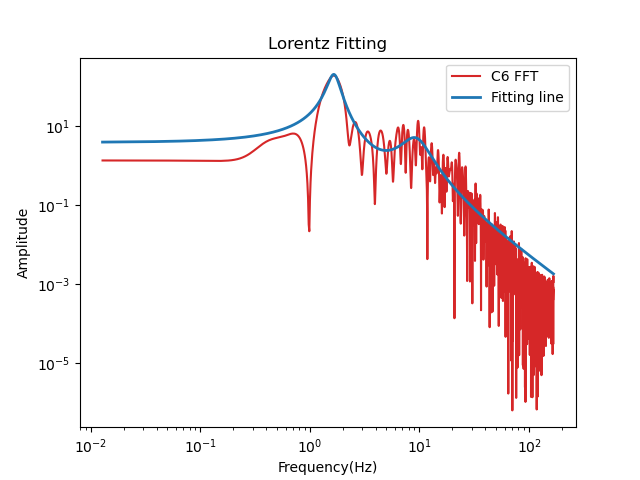}
        \caption{FFT of light curve}
        \label{fig:c6}
    \end{subfigure}
    \hfill
    \begin{subfigure}[b]{0.45\textwidth}
        \includegraphics[width=\textwidth]{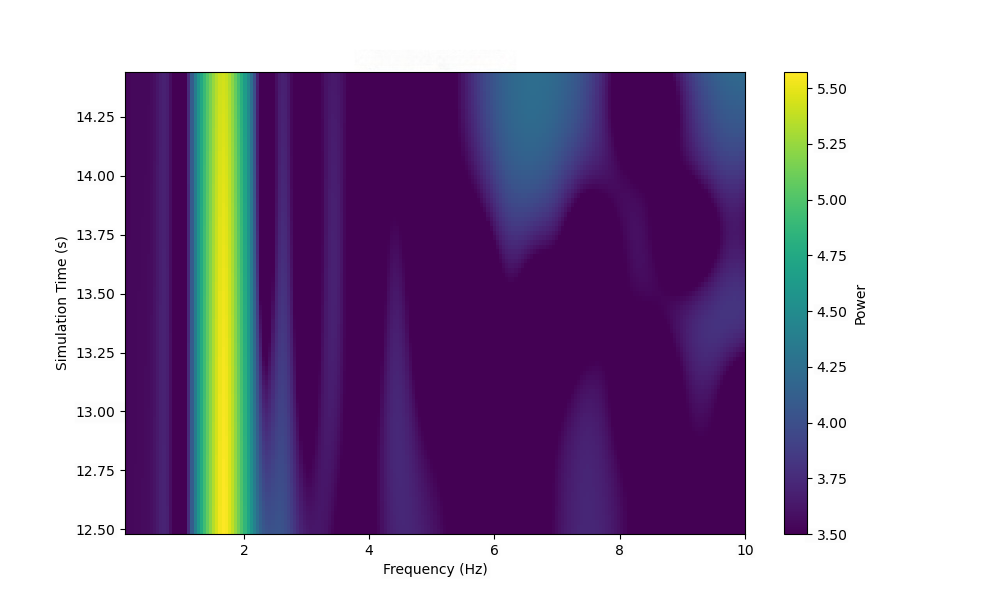}
        \caption{Weighted Wavelet Z-Transformation}
        \label{fig:c6w}
    \end{subfigure}
    \caption{Power density spectra of integrated total luminosity for Case 6. Figure(a): The FFT result shows the QPOs frequency at 1.65$Hz$ and 8.99$Hz$. Figure (b): The WWZ result shows the strongest periodical frequency at 1.67$Hz$.}
    \label{fig:c6two}
\end{figure}

\begin{figure}[ht]
    \centering
    \begin{subfigure}[b]{0.45\textwidth}
        \includegraphics[width=\textwidth]{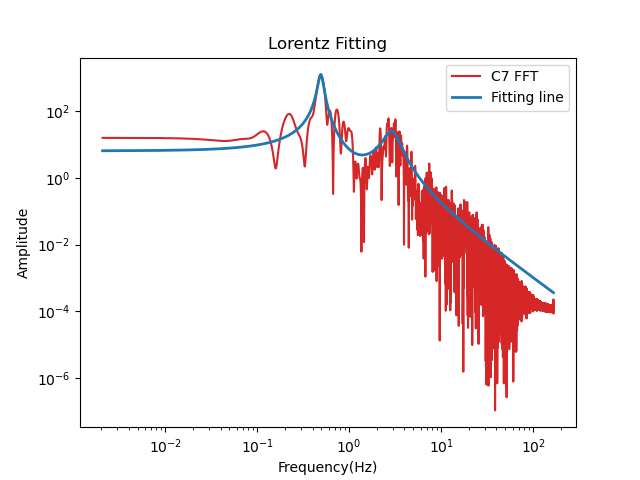}
        \caption{FFT of light curve}
        \label{fig:c7}
    \end{subfigure}
    \hfill
    \begin{subfigure}[b]{0.45\textwidth}
        \includegraphics[width=\textwidth]{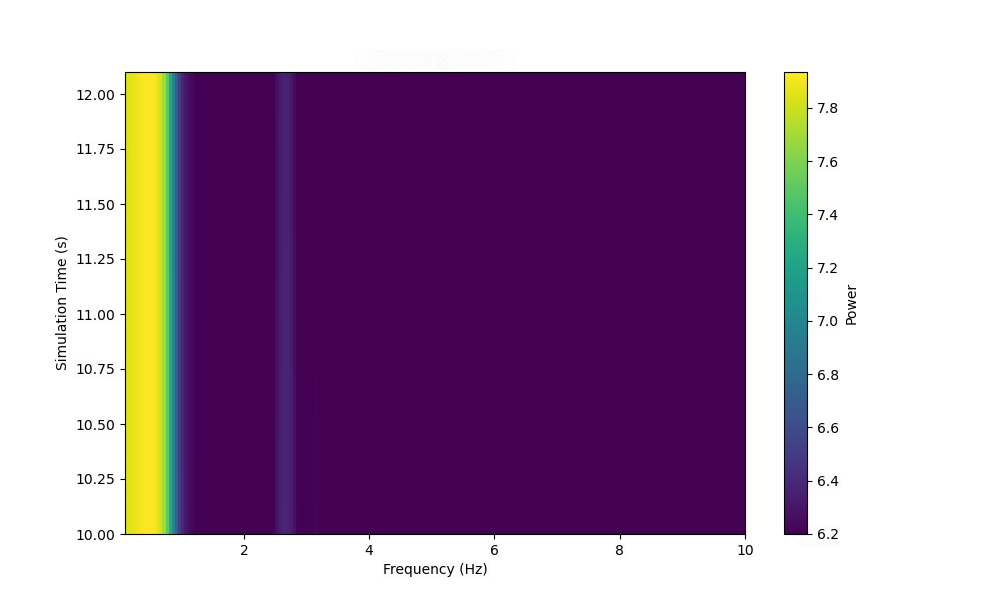}
        \caption{Weighted Wavelet Z-Transformation}
        \label{fig:c7w}
    \end{subfigure}
    \caption{Power density spectra of integrated total luminosity for Case 7. Figure (a): The FFT result shows the QPOs frequency at 0.49$Hz$ and 2.9$Hz$. Figure (b): The WWZ result shows the strongest periodical frequency at 0.44$Hz$.}
    \label{fig:c7two}
\end{figure}

The combined FFT and WWZ analyses reveal coherent yet temporally evolving QPO signatures across our simulation cases. In Case 1 (Fig. \ref{fig:c1two}), dual FFT peaks at 40.17$Hz$ (Q=50.28) and 146.57$Hz$ (Q=203.58) coexist with a WWZ-identified dominant signal at 146.67$Hz$ localized within 10.2–11.2$s$, establishing the higher-frequency component as the primary oscillation mode. Case 3 (Fig. \ref{fig:c3two}) exhibits bifurcated frequencies through FFT detection of 8.15$Hz$ (Q=7.69) and 28.89 Hz (Q=9.26), while WWZ analysis isolates a transient 7.85$Hz$ signal during 12.0–13.5$s$, indicative of dynamic mode switching. For Case 5 (Fig. \ref{fig:c5two}), the FFT-resolved 1.99$Hz$ (Q=2.21) and 8.61$Hz$ (Q=1.68) components contrast with WWZ's sustained 2.28$Hz$ oscillation spanning 4.0–8.5$s$, demonstrating temporal decoherence in higher-frequency modes. Similarly, Case 6 (Fig. \ref{fig:c6two}) shows FFT peaks at 1.65$Hz$ (Q=3.75) and 8.99$Hz$ (Q=1.66), whereas WWZ detects a stable 1.67$Hz$ signal during 12.5–14.5$s$, correlating with the stronger low-Q component. The low-frequency regime in Case 7 (Fig. \ref{fig:c7two}) presents FFT-identified 0.49$Hz$ (Q=7.66) and 2.90$Hz$ (Q=2.50) peaks, consistent with WWZ's baseline 0.44$Hz$ fluctuation. Crucially, Table 2 highlights Cases 5–6 where secondary peaks (8.61$Hz$/Q=1.68 and 8.99$Hz$/Q=1.66) exhibit $Q \le 2$, justifying their attenuated WWZ signatures compared to primary components — a quantitative confirmation of Q-value's efficacy in distinguishing physically significant oscillations from broadband noise.

Recently, \citet{2024MNRAS.531.1149N} reported that they employed the LAXPC (Large Area X-ray Proportional Counter) instrument, to observe the black hole X-ray binary Swift J1727.8-1613. Additional data was obtained from the NICER (Neutron Star Interior Composition Explorer), the MAXI (Monitor of All-sky X-ray Image) instrument, and the Swift/BAT (Burst Alert Telescope) to find the LFQPOs in hard X-rays up to 100 keV during the unusual outburst phase of Swift J1727.8-1613. They detected the energy range of 20-100 keV, the observed LFQPO has a centroid frequency of 1.43$Hz$ and a coherence factor Q of 7.14. In their calculation of the Q value, there is a missing factor of 1/2 compared to ours. When this factor of 1/2 is included, the result is very close to our Case 6.

Moreover, according to \citet{2024RAA....24c5001K}, he discovered a QPO with a frequency of 7.70 Hz through detailed analysis of the observational data of Swift J1728.9-3613. This discovery was made using data collected by NICER. The finding showed a slight discrepancy compared to the FFT results for Case 3 but was highly consistent with the WWZ analysis results. According to \citet{2024MNRAS.532.4486A}, they also observed type-B and type-C QPOs with frequencies ranging from 0.10 to 5.37 $Hz$, along with their harmonics in the source GX 339-4. For H1743-322, prominent type-C QPOs were detected in the frequency range of 0.22 to 1.01 $Hz$, accompanied by distinct harmonics. These observations were made using data from the AstroSat and NuSTAR(Nuclear Spectroscopic Telescope Array) archives, along with additional data from the MAXI/GSC and Swift/BAT light curves. These discoveries are in good agreement with our simulation results.

\begin{figure}[htbp]
    \centering
    \includegraphics[width=.45\textwidth]{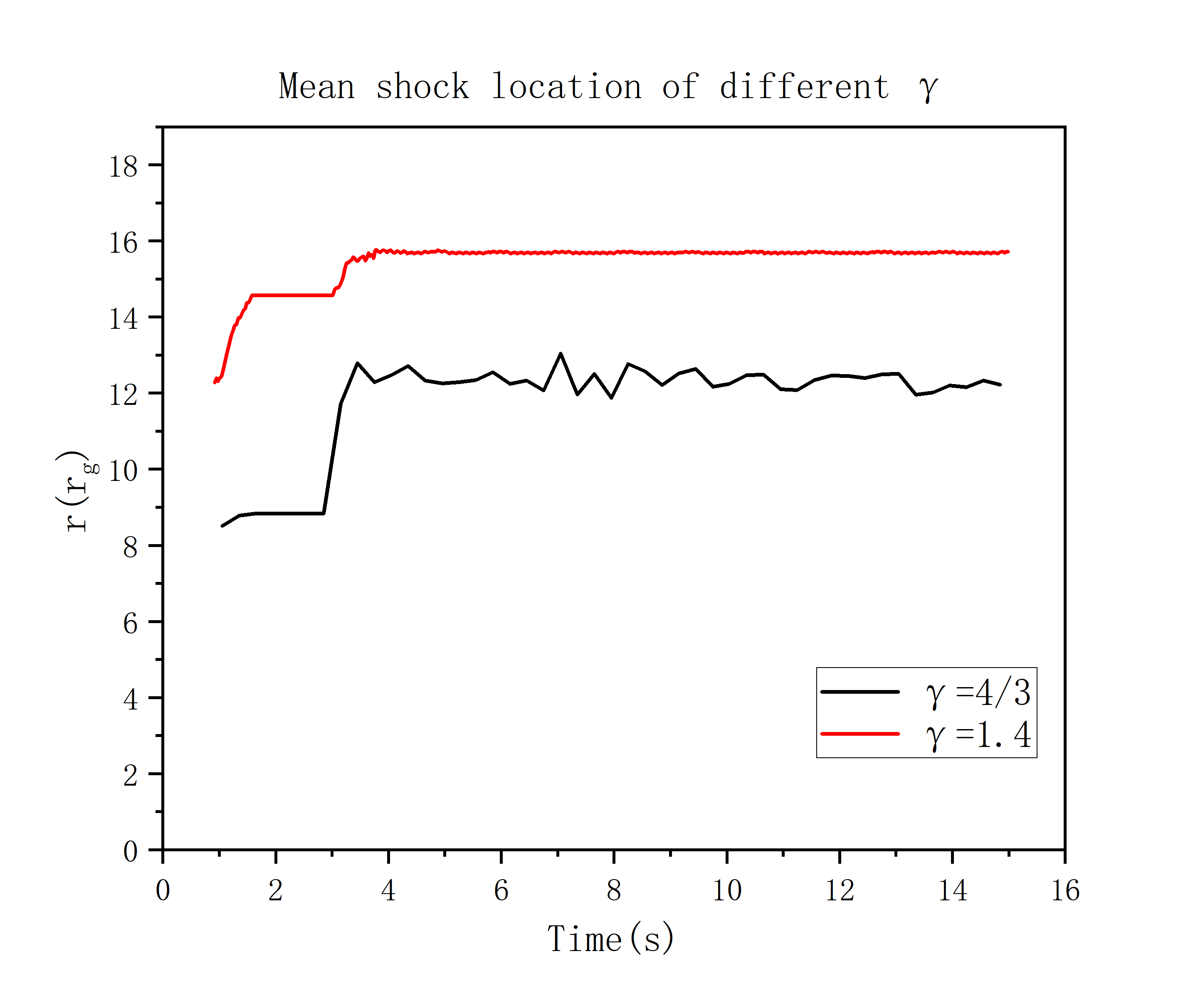}
    \caption{Shock location versus time for a particular case with the same specific angular momentum and energy and different $\gamma$.}
    \label{fig:gamma}
\end{figure}

Besides, we considered the effects of the different values of $\gamma$ in the system. In the work by \cite{2002ApJ...577..880D}, it was shown that varying $\gamma$ has a significant impact on the accretion flow dynamics. Our results agree with them, we took $\gamma =5/3, 1.5, 1.4$ and $4/3$, and when $\gamma \to 5/3$ it is the non-relativistic case, we deduced the analytical results. For $\gamma = 1.5$, the sonic point is available, and in some simulation setups, the shock doesn't exist. So we chose the $\gamma = 1.4$ and made a comparison with $\gamma = 4/3$ in Fig.\ref{fig:gamma}. The most obvious difference is the shock location. When the $\gamma$ increases, the shock location moves further from the BH. 

\begin{figure}[htbp]
    \centering
    \begin{subfigure}[b]{0.45\textwidth}
        \includegraphics[width=\textwidth]{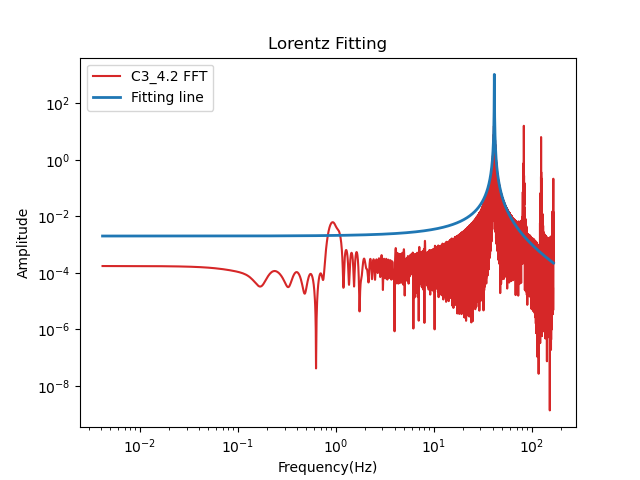}
        \caption{FFT of light curve}
        \label{fig:c3_4.2}
    \end{subfigure}
    \hfill
    \begin{subfigure}[b]{0.45\textwidth}
        \includegraphics[width=\textwidth]{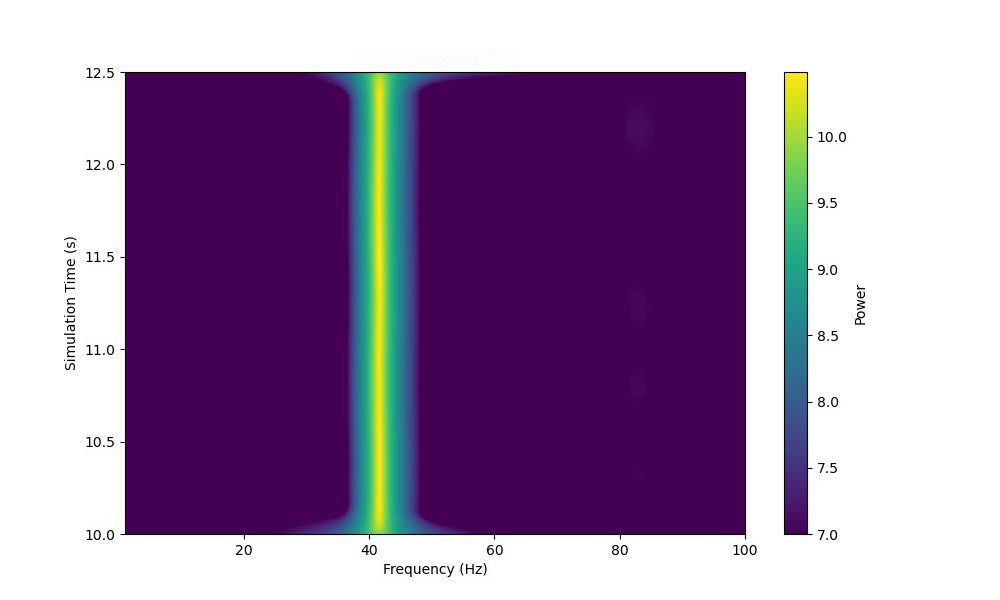}
        \caption{Weighted Wavelet Z-Transformation}
        \label{fig:c3_4.2w}
    \end{subfigure}
    \caption{Power density spectra of integrated total luminosity for Case 3 with $\gamma = 1.4$. Figure (a): The FFT result shows the QPOs frequency at 41.56$Hz$. Figure (b): The WWZ result shows the strongest periodical frequency at 41.67$Hz$.}
    \label{fig:c3_4.2two}
\end{figure}

We took the same approach, including FFT and WWZ to analyze the light curve. As shown in Fig.\ref{fig:c3_4.2two}, there is still a QPOs phenomenon, with the frequency of 41.56$Hz$, which is higher than that in the same case with the same specific energy and angular momentum $\gamma=4/3$. The Q value is 364.77. In the Lorentzian fitting curve, there is only one peak.

\section{Summary and conclusions}
%%\label{}
In this work, we performed a set of simulations with different initial parameters for the advective accretion flows around BHs, namely, specific energy and specific angular momentum. As the specific angular momentum of the accretion flow increases, the shock structure is harder to sustain, and the shock location moves further away from the black hole. 

After the formation of stable shocks in those flows, we introduce nonaxisymmetric perturbations in the system. During further evolution, QPOs were observed in several cases. The distribution of QPO frequencies ranges from 0.44 to 146.67 Hz. Our results are consistent with several observational signatures in black hole X-ray binaries (BHXRBs).

We also found that with a change in the adiabatic index, the flow behaviour changes correspond to different shock locations and associated QPOs.

Compared to previous simulation works by \citet{1999ApJ...516..411M}, we used a more sophisticated numerical tool, the PLUTO code, which is a high-resolution shock-capturing code. We performed our simulation runs in different computational domains of a larger size to extract physically meaningful results. Besides, we analyzed the data using two different methods (FFT and WWZ), both of which have given a range of QPO frequencies. The detailed comparison with observational signatures was presented for different stellar-mass black hole sources which was not done in earlier works, e.g. by \citet{1999ApJ...516..411M}.
We would like to add that similar simulation works but of non-spherical low angular momentum accretion flows around black holes with different initial conditions have been explored by $D\ddot{o}nmez$ and collaborators in case of Bondi-Hoyle-Lyttleton accretions in general relativistic HD and concluded that the evolution of flip-flop instabilities of the shock cone structures around non-rotating and rotating black holes are responsible for the physical origin of LFQPOs as well as HFQPOs (\citealt{2011MNRAS.412.1659D}, \citealt{donmez2025evolutionshockstructuresqpos}, and reference therein).

%First of all, when introducing the nonaxisymmetric perturbation to an accretion disk with a stable shock structure, it can be deduced QPOs. Moreover, QPOs are more difficult to detect in simulations with high angular momentum and low specific energy. 

%Using different adiabatic indices, it led to significant differences in the results.

\section*{Acknowledgements}
This work is supported by the National Natural Science Foundation of China under grant No. 12073021. The authors thank Mr. Yuanshang Huang and Dr. Kaushik Chatterjee of SWIFAR, Yunnan University for their helpful discussions and suggestions regarding this work. The numerical simulations were conducted on the Yunnan University Astronomy Supercomputer (YUNAS) and analyzed by the open-source tool ShockFinder (see https://www.github.com/wacmkxiaoyi/shockfinder).

\appendix

%\section{Appendix title 1}

%\section{Appendix title 2}

\bibliographystyle{elsarticle-harv} 
\bibliography{main}



%% For example:

%% \bibitem[Aladro et al.(2015)]{Aladro15} Aladro, R., Martín, S., Riquelme, D., et al. 2015, \aas, 579, A101

%%\end{thebibliography}

\end{document}